\documentclass{aa}
 \usepackage{amssymb}

\usepackage[colorlinks, citecolor=blue]{hyperref}
\usepackage{multirow}
\usepackage{rotating}
\usepackage{ctable}
\usepackage{array}
\usepackage{hyperref}

\usepackage{graphicx,epsfig,fancyhdr,psfig,rotating,amsmath,epsf,txfonts,natbib,epstopdf,multirow}
\def \kms {{ \rm km\;s$^{-1}$}}

\def \lm {}

\begin{document}

\title{A coronal wave and an asymmetric eruptive filament in SUMER, CDS, EIT, and TRACE co-observations}

\author{M.S. Madjarska, J.G. Doyle, J.  Shetye}

\offprints{madj@arm.ac.uk}
\institute{Armagh Observatory, College Hill, Armagh BT61 9DG, N. Ireland}
 \date{Received date, accepted date}

\abstract
{The investigation covers the complex subject of coronal waves and the phenomena contributing to and/or causing their formation. }
{The objectives of the present study is to provide a better physical understanding of the complex  inter-relation and evolution of several solar coronal features comprising a double-peak  flare, a coronal dimming caused by a Coronal Mass Ejection (CME), a CME-driven compression, and a fast-mode wave. For the first time, the evolution of an asymmetric eruptive filament is analysed in simultaneous Solar Ultraviolet Measurement of Emitted Radiation  (SUMER) spectroscopic and Transition Region and Coronal Explorer (TRACE) and Extreme-ultraviolet  Imaging Telescope (EIT) imaging data.}
{We use imaging observations from EIT and TRACE in the 195~\AA\ channel and spectroscopic observations from the Coronal Diagnostic Spectrometer (CDS) in a rastering and SUMER in a sit-and-stare observing mode.  The SUMER spectra cover spectral lines with formation temperatures from $logT(K)$ $\sim$ 4.0  to 6.1.}
{Although the event was  already analysed in two previous studies, our analysis brings a wealth of new information on the dynamics and physical properties of the observed phenomena. We found that the dynamic event is related to a complex flare with two distinct impulsive peaks, one according to the Geostationary Operational Environmental Satellite (GOES) classification as C1.1 and the second -- C1.9. The first energy release triggers a fast-mode wave and a CME with a clear CME driven compression ahead of it.  This  activity is related to, or possibly caused, by an asymmetric filament eruption.  The filament is observed to rise with  its  leading edge moving at a speed of $\sim$300~\kms\ detected both in the SUMER and CDS data. The rest of the filament body moves at only $\sim$150~\kms\  while untwisting.  No signature is found of the fast-mode wave in the SUMER data, suggesting that the plasma disturbed by the wave had temperatures above 600\,000 K. The erupting filament material is found to emit only in spectral lines at transition region temperatures. Earlier identification of a coronal response detected in the Mg~{\sc x}~609.79~\AA\ line is  found to be caused by a blend from the O~{\sc iv}~609.83~\AA\ line. }
{We present a unique analysis of the complex phenomenon called `EIT/Coronal Wave', confirming its bimodal nature. We suggest that the disintegration of the dimming/CME and the CME-driven compression are either caused by a CME-CME interaction taking place in the low solar atmosphere or by an  impulsive CME cavity overexpansion in the low solar atmosphere.}
 
\authorrunning{Madjarska, M. S. et al.}
\titlerunning{A coronal wave \& an asymmetric filament eruption}

\maketitle

\section{Introduction} 
\begin{table}[ht!]
\centering
\caption{SUMER spectral lines. The expression `/2' denotes that the spectral
 line was observed in second order and `b' that the line is blended by a close-by line.}
\begin{tabular}{l l c}
\hline\hline
Ion & Wavelength (\AA) & logT$_{max}$ (K)\\
\hline
S~{\sc iv} &	750.22		&			5.0\\
O ~{\sc v}	&	758.68		&			5.4\\
O ~{\sc v} 		&		759.43b 			& 5.4\\		
S ~{\sc iv} &  759.34b 	 &  5.0\\										
O ~{\sc v} 		&				762.43		&			5.4\\	
O ~{\sc v} 		&				760.43		&			5.4\\	
N ~{\sc iii} 	&					762.00	&				 4.9\\
N ~{\sc iii}  &						764.36					&4.9\\
N ~{\sc iv} 	&					765.15		&			5.1\\
Ne ~{\sc viii}  &						770.42		&			5.8\\
Ne ~{\sc viii}  &						780.30		&			5.8\\
S ~{\sc  v } 				&		786.47		&			5.2\\
O ~{\sc iv } 				&		787.72		&			5.2\\
O ~{\sc  iv } &	790.11b		& 5.2\\
O ~{\sc iv } & 790.19b &5.2\\
N~{\sc v} & 1238.82 & 5.3 \\
Fe ~{\sc xii}	&	1241.95	&		6.1\\
N ~{\sc v}	&	1242.80		&	5.3\\		
C ~{\sc i}		& 1243.52			& 4.0\\ 	
C ~{\sc i}	&	1244.00		&	4.0\\	
C ~{\sc i}	&	1244.51		&	4.0\\
C ~{\sc i}	&	1244.99		&	4.0\\
C ~{\sc i}	&	1245.18		&	4.0\\
C ~{\sc i}	&	1245.53		&	4.0\\
C ~{\sc i}	&	1245.94		&	4.0\\
C ~{\sc i}	&	1246.17		&	4.0\\
C ~{\sc i}	&	1246.87		&	4.0\\
S~{\sc i}   &       1247.16           &      4.0\\
C ~{\sc i}	&	1247.86		&	4.0\\	
C ~{\sc i}	&	1248.00		&	4.0 \\
Mg~{\sc x}/2 & 1249.90b & 6.1 \\
Si~{\sc ii} & 1250.09b&4.1   \\
O~{\sc v}/2&1259.54&5.4  \\
\hline
\end{tabular}
\label{table1}
\end{table}

\label{intro}
Coronal waves  (CWs), also known as Extreme-Ultra-Violet (EUV) and Extreme-ultraviolet  Imaging Telescope (EIT) waves (hereafter coronal waves), are large-scale coronal transients observed in the form of  a diffuse brightening that were first detected in  observations  taken by EIT \citep{1995SoPh..162..291D} on board the Solar and Heliospheric Observatory \citep[SoHO,][]{1998GeoRL..25.2465T}. These waves are usually observed as moving, bright quasi-circular rings of increased emission in EUV wavelengths, mainly in the 195~\AA\  imaging channels (1.6 MK) with a typical velocity of 200--400\kms.   In 171~\AA\  channels the CWs are seen as emission reduction with respect to the surrounding coronal emission, which is  caused by heating and thus the ionization of  Fe{\sc~ix/x}  to higher ionisation states.  Coronal waves are caused by the shock produced from the sudden energy release in active regions and are associated with CMEs rather than flares  \citep{2002ApJ...569.1009B}. Not all CMEs, however,  are found to generate coronal waves \citep{2002ApJ...569.1009B}. Moreton waves (i.e. a chromospheric wave, first reported by \citet{1960AJ.....65U.494M} in H$\alpha$ observations propagating with 1000~\kms)  are believed to be the chromospheric  counterpart of coronal waves,   with both waves showing similar behaviour \citep[][and the references therein]{2006ApJ...647.1466V}.  \citet{2012ApJ...745L..18A} reported on the first simultaneous co-spatial detection of a Moreton  and a coronal wave. 

Thanks to the high-cadence images from the Atmospheric Imaging Assembly (AIA) on board the Solar Dynamic Observatory (SDO),  it was found that CWs show deceleration which is a typical characteristic of large-amplitude waves   \citep{2004A&A...418.1101W, 2008ApJ...680L..81L, 2008ApJ...681L.113V}. Coronal wave fronts  are reported to be anisotropic and non-homogeneous \citep{1999ApJ...517L.151T}. From observations with the Extreme Ultraviolet Imager (EUVI)  on board STEREO-B,   \citet{2010ApJ...716L..57V} found that coronal waves are dome shaped with erupting CME loops inside the dome. The CW properties can be found tabulated in \citet{2011SSRv..158..365G}.

The nature of the phenomenon `coronal wave' is strongly  disputed. The interpretations include fast-mode waves \citep{1999SoPh..190..467W, 2001ApJ...560L.105W, 2010ApJ...713.1008S}, slow-mode waves or solitons \citep{2007ApJ...664..556W}, and non-waves related
to a current shell \citep{2008SoPh..247..123D} or successively reconnecting magnetic field lines at
a CME front \citep[][and the references therein]{2014SoPh..289.2565D, 2010ApJ...718..494A}. More details on the properties and modelling of coronal waves can be found in several reviews by \citet{2008SoPh..253..215V}, \citet{2011SSRv..158..365G}, and  \citet{2012SoPh..281..187P}. The most recent review by \citet{2014arXiv1404.0670L} summarises the very latest findings from AIA/SDO observations and  modelling of coronal/EIT waves. The authors conclude that CWs clearly have a hybrid or bimodal nature with an outer EUV front of a true fast-mode wave that  travels ahead of an inner non-wave
component of CME-driven compression (see the composed Figure~2 in their paper). They also suggest that heating due to electric current dissipation or magnetic reconnection may have a contribution to the EUV emission at the inner, CME front, but not the outer, true wave front. This complex nature of the phenomenon was already suggested  by \citet{2000JGR...10523153F}. \citet{2004A&A...427..705Z} analysed two events that initiated close to the limb permitting the investigation of both the wave and the structure of the CMEs. The authors concluded that the observations suggest a bimodal phenomenon where the wave mode represents a wave-like propagating disturbance observed as a bright front that propagates to large distances from the dimming sites and has a quasi-circular appearance. The propagation of a dimming and  EIT wave, as a result of successive opening of magnetic field lines
during the CME lift-off, is regarded as an eruptive mode.

Spectroscopic observations of coronal waves and related phenomena are very important as they provide valuable information on their physical properties but are very limited because of the low chance of encountering transient phenomena in their limited field-of-view (FOV). Until present only very few observations were reported based on data from  CDS on board SoHO and the EUV Imaging Spectrometer (EIS) on board Hinode \citep[e.g.][]{2003ApJ...587..429H, 2011ApJ...743L..10V, 2011ApJ...737L...4H, 2011ApJ...740..116C, 2013ApJ...775...39Y, 2013SoPh..288..567L}. 

\citet{2011ApJ...743L..10V} performed a detailed plasma diagnostics of a coronal wave based  on EIS/Hinode observations.  The authors measured Doppler velocities in Fe~{\sc xii} ($logT_{max}(K)$  = 6.1), Fe~{\sc xiii} ($logT_{max}(K)$  = 6.2),  and Fe~{\sc xvi} ($logT_{max}(K)$  = 6.4)  of up to 20~\kms\  (down-flows)  at the wave front followed by up-flows of up to 5~\kms. The down-flow represents the plasma pushed downwards by the wave followed by the up-flow, which is a signature of the relaxation of the plasma behind the wave front. Their density diagnostics reveals that enhanced density is not related to the wave front, but to the eruption behind  the wave, i.e. it is not due to plasma compression at the wave front itself. They found no signature of the wave in  He~{\sc ii} 256~\AA\ ($logT(K)$  = 4.7), which  indicates that the upper chromosphere was not affected by the wave. 

 \citet{2003ApJ...587..429H} first investigated a phenomenon associated with a coronal wave using  CDS raster observations. They studied the temperature response and the Doppler velocities of a feature trailing behind a coronal wave assumed to be an eruptive filament. The wave itself was not identified  in the spectroscopic data, but was analysed in detail in the TRACE co-observations first by \citet{1999SoPh..190..467W} and later by  \citet{2003ApJ...587..429H}. The event discussed in these two papers is the subject of the present study. 

\begin{figure}[!ht]
\centering
\includegraphics[scale=0.8]{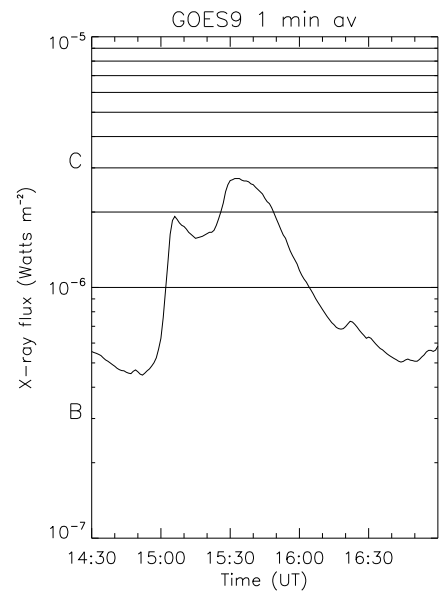}
\caption{GOES 9  1-min-averaged X-ray flux in the range 1 -- 8~\AA.}
\label{fig_goes}
\end{figure}

Here, we report the first study of a coronal wave and related phenomena including flare, coronal mass ejection and filament eruption, using SUMER 
observations. The coronal wave was already studied in TRACE data by  \citet{1999SoPh..190..467W}  and CDS data (including a trailing eruptive feature believed to be  a filament) by \citet{2003ApJ...587..429H} (hereafter WT and HS, respectively). In our study we combined temporally and spatially the information provided by all three instruments, i.e. CDS, TRACE, and SUMER.  We report the first SUMER  observation of an eruptive filament.  The paper is organised as follows: In Section~2, we describe the observational material. Section~3 introduces the studied  phenomena and details from the two earlier studies. The imaging and spectroscopic data analysis, results, and discussion are given in Section~3.1 and 3.2, respectively. The  conclusions  are outlined in Section~4.

\section{Observations}
\label{sect2}

The observations used for the present study were taken on 1998 June 13 with the  Solar Ultraviolet Measurements of the Emitted Radiation (SUMER) spectrometer and the CDS  on board SoHO, and the TRACE imager. Synoptic observations in the H${\alpha}$ line from the Big Bear Solar Observatory were also used.   A coronal  wave was observed in the TRACE field-of-view followed by a dimming (the disk projection of a coronal mass ejection) originating in  NOAA 08237.  A flare with two impulsive phases, one reaching GOES C1.1 and the second reaching GOES C1.9  (Fig.~\ref{fig_goes}),  occurred in the time preceding the dimming and the wave. We will refer to this flare as `combined' flare.  The EIT, TRACE and CDS data are described in \citet{1999SoPh..190..467W} and \citet{2003ApJ...587..429H}. In the next section, we will provide some additional details on the CDS data that are important for the follow-up discussion.

The SUMER spectrometer  \citep{1995SoPh..162..189W, 1997SoPh..170...75W,1997SoPh..170..105L} was observing in a sit-and-stare mode with a 1\arcsec$\times$300\arcsec\ slit  and a 480~s exposure time on detector B. Two consecutive exposures in two spectral ranges centred at 770~\AA\ and 1238~\AA\ were taken in the disk centre starting at xcen=0\arcsec\ and ycen=0\arcsec\ at 15:39~UT. A compensation for the  solar differential rotational was applied during the observations. A full spectral window, i.e. 1024 spectral pixels or $\sim$40~\AA, was telemetered to the ground.  The observed spectral lines are listed in Table~\ref{table1}. Note that we list only the lines that were strong enough to be used in the present study.  Standard data reduction procedures were applied to the data. 

\begin{figure*}[!ht]
\centering
\includegraphics[scale=1.]{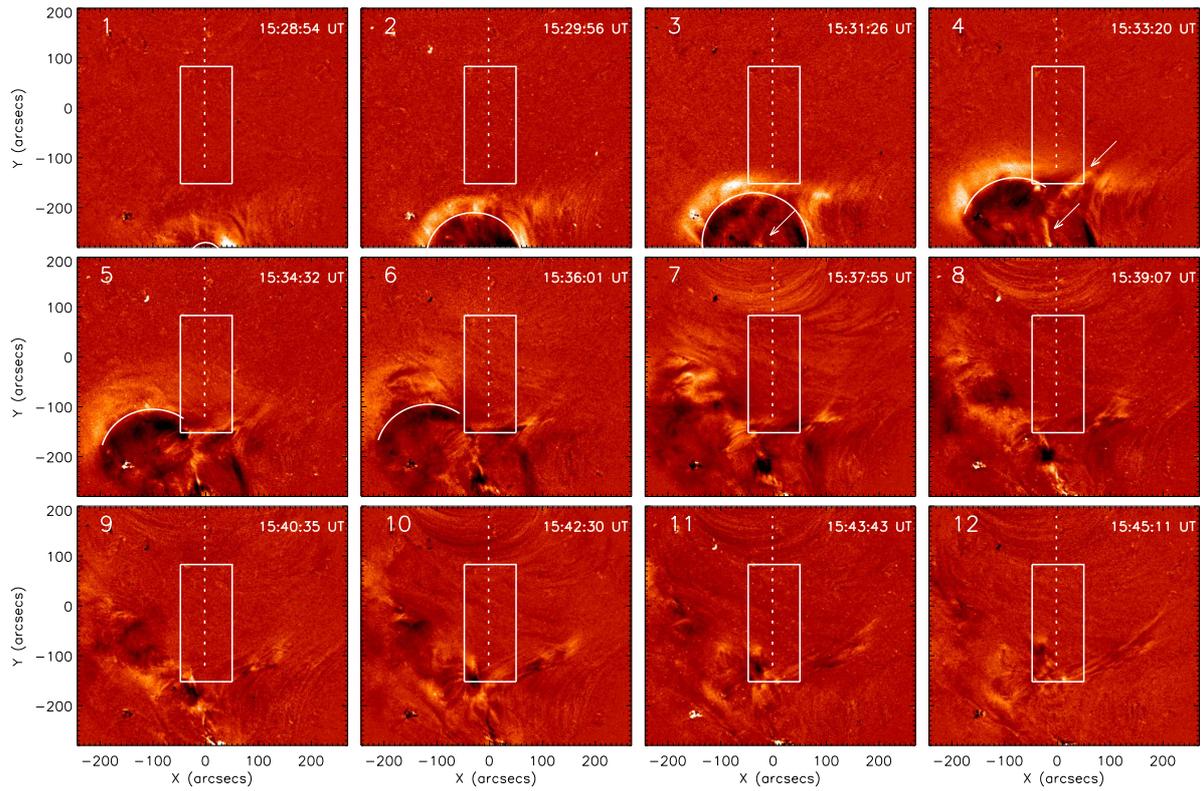}
\vspace{-1cm}
\caption{TRACE running difference images. The solid line rectangle represents the CDS FOV. The dashed vertical line is the SUMER slit position. The over-plotted semicircles outline the border between the CME-driven compression and the dimming/CME. }
\label{fig_trace}
\end{figure*}

\section{Analysis, results and discussion}

\subsection{Imaging analysis}

The way slit spectrometers operate often makes the registration of dynamic transient phenomena quite a lucky coincidence. Motivated by the fact that no SUMER observations of coronal waves have ever been reported, we investigated existing reports on coronal waves in the search of such data. We were very fortunate to discover that during SUMER  disk-centre observations, a `coronal wave' initiated in NOAA 08237 on 1998 June 13 propagated in a south-north direction moving under the SUMER slit. 
\begin{figure*}
\centering
\vspace{-2cm}
\includegraphics[scale=0.9]{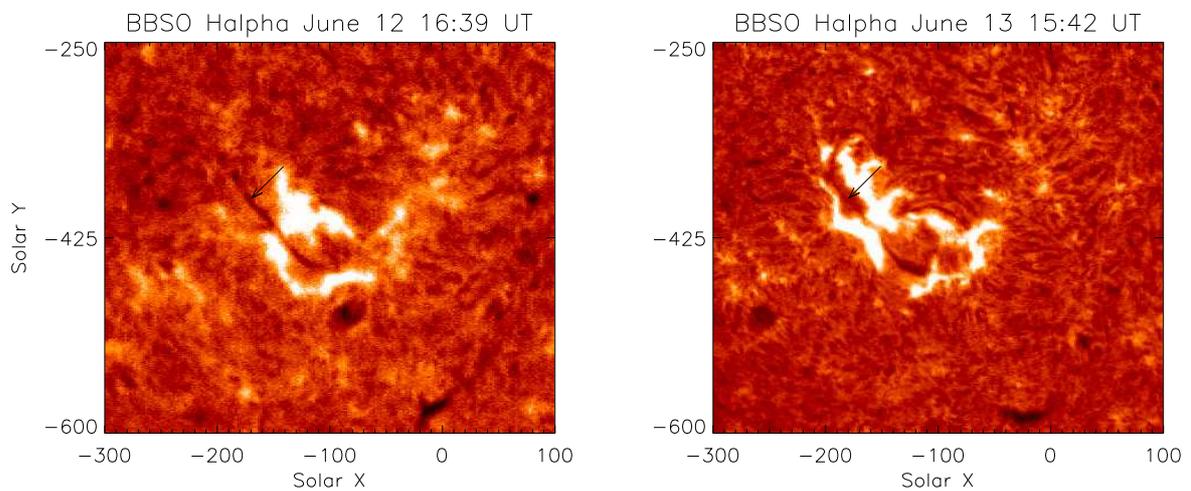}
\vspace{-2.5cm}
\caption{H$\alpha$ BBSO synoptic  images with a field-of-view covering NOAA 08237. {\bf Left:} A filament is visible lying along the magnetic neutral line. {\bf Right:}  The upper part of the filament  has disappeared leaving a small undisturbed part. The image is de-rotated to the time of the image taken on June 12.}
\label{fig_fil}
\end{figure*}

 The coronal phenomenon, shown here in Fig.~\ref{fig_trace}, is best seen in the difference (running or base) TRACE 195~\AA\ images, and appears as a bright ring ahead of a region with reduced emission. We will briefly review the findings of the two previous studies. First, \citet{1999SoPh..190..467W} described the coronal disturbance   as registered in the TRACE data and identified it as an `EIT wave'. From the absence of the feature in the  TRACE Lyman-$\alpha$  images and undisturbed low-lying loop structures (e.g. coronal bright points) with height up to 40\arcsec, the authors concluded that the wave travelled at a height above 15~Mm. The authors interpreted the phenomenon as a fast-mode, magneto-acoustic wave possibly driven by a CME.  From the combination of the 171~\AA\ and 195~\AA\ channels, they  estimated that the wave was propagating through plasma with temperature in the range  1--1.4~MK. Using Huygens plotting technique, the velocities of the wave front was  found to be in the range between 200 and 800~\kms. 
 
The second study by \citet{2003ApJ...587..429H}  also analysed  the phenomenon in the TRACE data but reached different conclusions.  They  identified a `weak' faster (500~\kms) and a `bright'  slower (200~\kms) wave (see their article for more details).  They also analysed  co-temporally taken CDS raster spectra and found that the `weak' wave moves through the CDS FOV, but no `substantial line-shifts' were detected in the CDS spectra that made the authors conclude that the velocity should be less than 10~\kms\ (i.e. below the detection capabilities of CDS). Further on, their study concentrates on a feature that moves through the CDS FOV following the wave. Their analysis suggested that this phenomenon could possibly be a filament  that had erupted during the flare. The Doppler-shift analysis showed two velocity components: one at 150~\kms\  in the lower part of the CDS FOV and a second component of 350~\kms\ that `proceeds from south to north'.  The authors note that `the faster (350~\kms) component is moving at a speed around 200~\kms\ in a north or northeasterly direction'.  The phenomenon was detected  in the Mg~{\sc x} \lm609.79~\AA, O~{\sc v} \lm629.73~\AA, and He~{\sc i} \lm584.33~\AA\  lines, and the authors concluded that the material represents a cool filament ejected together with its hot surrounding.  \citet{2003ApJ...587..429H} reached the conclusion that the observations show similarities with the numerical  model of \citet{2002ApJ...572L..99C} where a piston-driven shock (the `weak faster wave') is triggered by a rising flux rope (i.e. the erupting filament). The `bright wave' ,  which has a slower propagation ($\approx$200\kms), represents  successive opening of the magnetic field lines covering the flux-rope/filament. Both WT and HS suggest that the event is driven by a CME.

\begin{table}
\centering
\caption{Dimming velocities measured in the TRACE running-difference images shown in Fig.~\ref{fig_trace}.}
\begin{tabular}{c c c}
\hline
\multicolumn{2}{c}{Time (UT)} & {Velocity (km/s) } \\
From & To & $\pm$20\kms\\
\hline
15:29:56 &15:31:26 & 240 \\ 
15:31:26 & 15:33:20 & 230\\ 
15:33:20 & 15:34:32 & 220\\ 
15:34:32 & 15.36:01 &  80\\
\hline
\end{tabular}
\label{table2}
\end{table}

 To correctly interpret  the SUMER observations, we started our analysis by re-examining the TRACE and CDS data and combining them with the SUMER observations. In Fig.~\ref{fig_trace}, we show a sequence of running difference images in the TRACE~195~\AA\ channel over-plotted  with  the CDS raster field-of-view (FOV) and SUMER slit position.  We would like to note that this study does not aim to help resolve the debate about the nature of the phenomenon `coronal/EIT/EUV wave'. We, however, strongly believe that the past ten years of theoretical and observational studies have provided enough evidence supporting the interpretation of coronal waves  as  having a hybrid nature, i.e an outer EUV front of a true fast-mode wave that  travels ahead of an inner non-wave component of  a CME-driven compression (hereafter CDC)    \citep[e.g.][]{2012ApJ...754....7S, 2012ApJ...750..134D}. The wave and non-wave feature (CDC) are often hard to distinguish in certain coronal conditions or when the propagation is close to the source region \citep{2011ApJ...728....2D}.

\begin{figure*}[!ht]
\centering
\includegraphics[scale=1.]{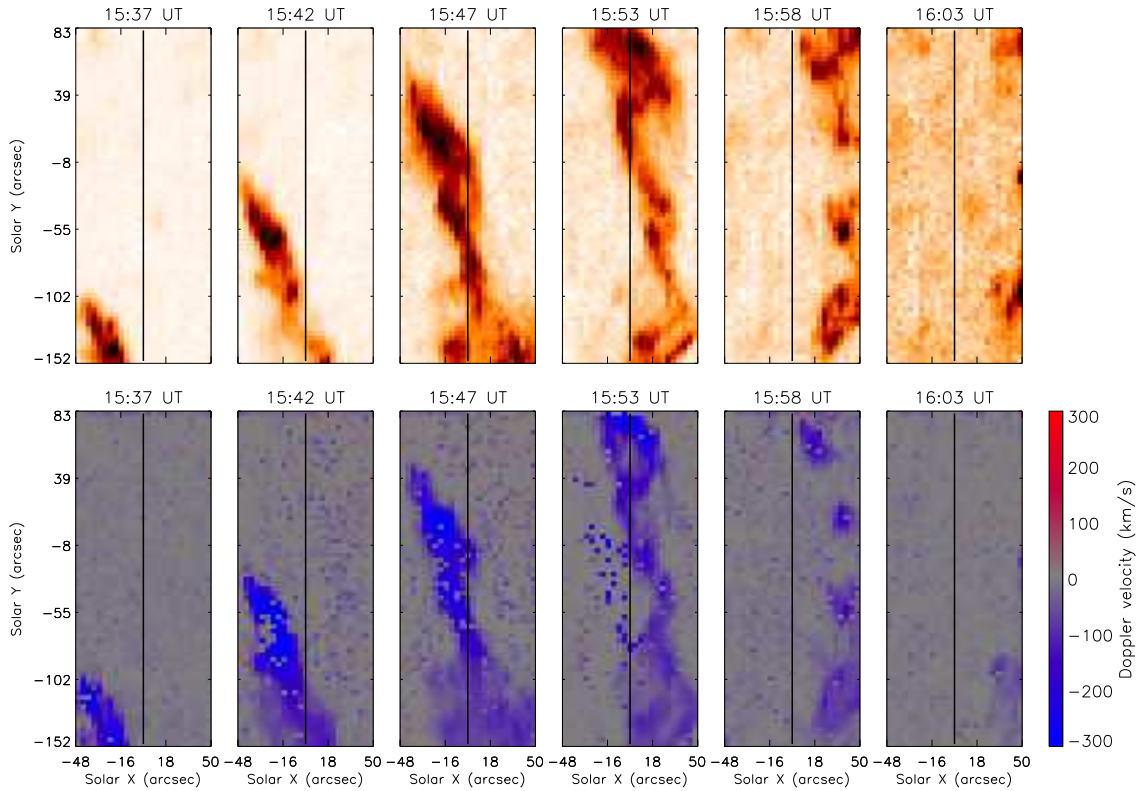}
\vspace{-1cm}
\caption{{\textbf CDS O~{\sc v}~629.74~\AA\ radiance and Doppler-shift images showing the eruptive filament.}}
\label{fig_vel_cds}
\end{figure*}

In light of the above definition, the  `weak' fast wave as described by  \citet{2003ApJ...587..429H} clearly represents the true fast-mode  wave moving ahead of the `bright wave', which is  the CME-driven compression  \citep{2012ApJ...750..134D, 2011ApJ...728....2D}, also called a plasma pileup \citep{2000JGR...10523153F}.  After entering  the TRACE FOV (at 15:28~UT, panel 1 in Fig.~\ref{fig_trace})  the true wave  and the CME-driven compression are impossible to separate as they are still very close to the source region, i.e. the active region.  After 15:33~UT the wave moves ahead while the CDC disperses following the disintegration of the   dimming/CME  (discussed  later in the text). 

 As we mentioned above, the flare presents two phases of energy release, the first  at 15:06~UT classified as GOES C1.1 and the second at 15:34~UT classified as GOES C1.9. We estimated that the fast-mode wave driven by a CME was generated shortly  after the first energy release. We used the available EIT data (see the online material Fig.~\ref{fig_movie}) to obtain the position of the wave front as early as possible after the flare occurred  (i.e. 15:24~UT) and  to compare it with the wave front best seen in TRACE  at 15:29~UT (Fig.~\ref{fig_trace}). Note that the wave-front velocity could not be investigated in great detail because of the time cadence of the EIT and TRACE images.  The wave-front velocity was found to be 500$\pm$20~\kms\ calculated from both the EIT and TRACE images. 

The sharp transition between the dimming (created by the CME bubble passing through the corona) and the bright semicircle (the CDC \& the wave) outlined  in Fig.~\ref{fig_trace}, approximately  traces the edge of the CME bubble.  In  Fig.~\ref{fig_trace}, we present a series of TRACE 195~\AA\  running difference images showing the series of events that took place.  In the online material (Fig.~\ref{fig_movie}), we provide an animated sequence of images including, EIT running difference, TRACE~195~\AA\ original, and TRACE running difference images, temporally combined. The tip of the dimming first appears in the TRACE FOV at 15:28~UT (panel 1 in Fig.~\ref{fig_trace}). To separate the dimming expansion from its lateral displacement, we fitted a semicircle of a certain diameter on the sharp boundary between the dimming and the CDC. Between panels 1 and 2 the semicircle diameter increased from 30\arcsec\ (22~Mm) to 90\arcsec\ (65~Mm). As discussed in WT, the expected propagation height  of the CW  at this time is above 15~Mm. Low-lying coronal structures, such as coronal bright points, remained undisturbed by the wave (bright points heights range from 5 to 20~Mm \citep{2004A&A...414..707B, 2007AdSpR..39.1853T}. 
The lateral speed of the dimming  is  shown in Table~\ref{table2}. Note that this is not the actual rising speed of the CME, but it  represents  the  disk projection of a feature rising above the solar surface in a non-radial direction. This speed can be taken as a very low limit of the CME rise speed. For instance,  \citet{2010ApJ...716L..57V} found that  the velocity derived for the upward expansion of a wave dome above the limb  was  650~\kms,  while  the lateral expansion of the wave observed on-disk was 280~\kms.
 
 From 15:29~UT to 15:34~UT, the dimming moves with a speed of $\approx$230$\pm$20~\kms. After 15:34~UT, its speed sharply  drops to 80~\kms\ (see Table~\ref{table2}). The last image where the dimming is distinguishable enough to permit us to fit a semicircle is at 15:36~UT (panel 6 in Fig.~\ref{fig_trace}).  During the next three mins the whole dimming and the CDC  region are fully dispersed. The true wave, however, continues  moving undisturbed  across the solar disk  (northwards), reaching large coronal loops connecting NOAAs 08239 and 08238 to the north,  triggering their excitation and 
resulting oscillation. The fast wave separates from the dimming/CME  after 15:33~UT. The loop oscillations are  clearly seen in panel 7 at 15:37~UT in Fig.~\ref{fig_trace} but it can already be spotted in panel 6 at 15:36~UT.
 
 The time (15:33~UT) after which the CDC is seen `breaking' coincides with the appearance of a fast-moving bright elongated feature. The tip of this feature can be first noticed at the centre-bottom  part of the dimming region at 15:31~UT (white arrow at 15:31~UT in the bright wave in Fig.~\ref{fig_trace}). On the next image, at 15:33~UT, the feature has  made a south to north-east motion with a speed of $\approx$320$\pm$10~\kms. Another similar looking feature (not visible at 15:31~UT) is seen above  `breaking'  through the dimming and the  CDC region (see the upper arrow in  Fig.~\ref{fig_trace}). At this time, the fast-mode wave is clearly seen moving ahead.

\begin{figure*}[!ht]
\centering
\includegraphics[scale=0.9]{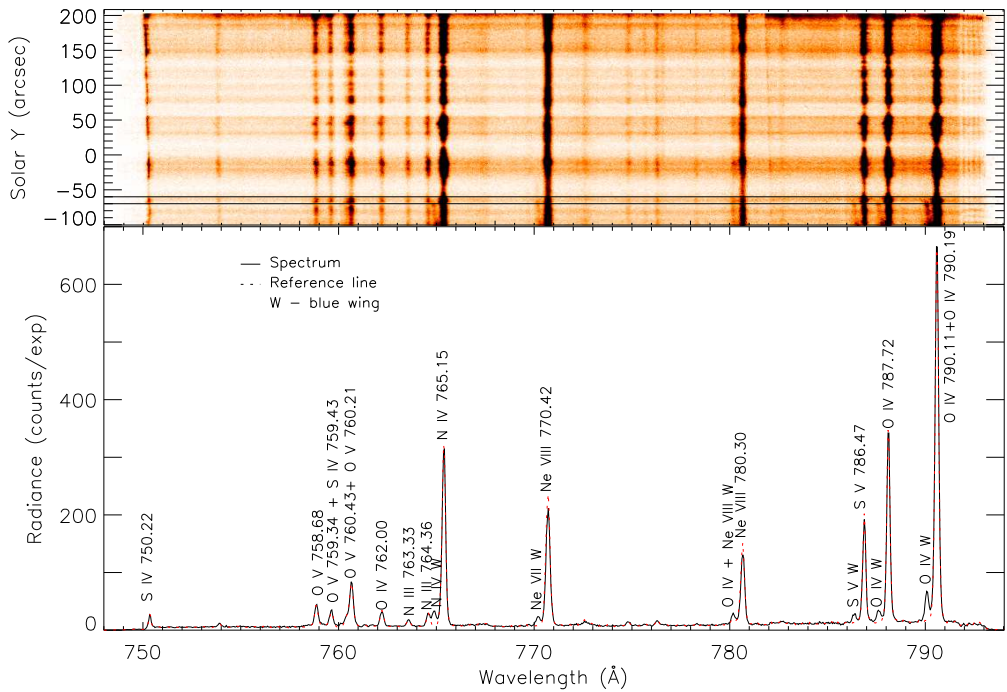}
\vspace{-1cm}
\caption{{\bf Top panel:} SUMER detector image taken at 15:39~UT. The region between the horizontal solid lines was used to produce the spectrum during the event. The same region was taken from the spectrum obtained earlier at 15:31~UT (before the event) to produce the reference spectrum. {\bf Bottom panel:} The spectrum during the event  is shown with a solid line, while the reference spectrum is shown with a dotted (red in the online version of the paper) line. `W' next to the line, e.g. `Ne~{\sc viii}~W',  denotes the blue wing of the line. }
\label{fig_sumer1}
\end{figure*}

\begin{figure*}[!ht]
\centering
\includegraphics[scale=0.9]{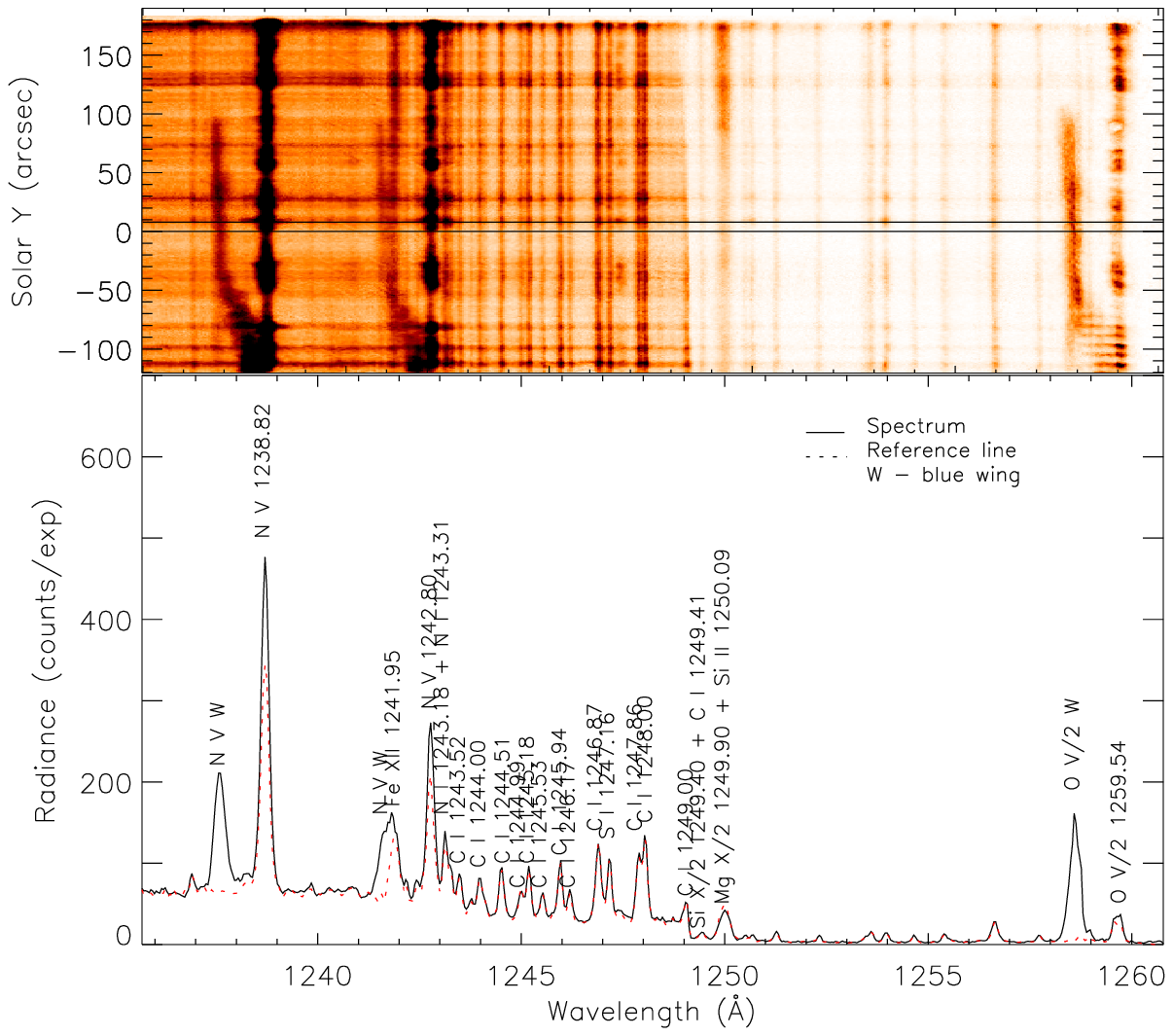}
\vspace{-1cm}
\caption{{\bf Top panel:} SUMER detector image taken at 15:48~UT. The region between the horizontal solid lines was used to produce the spectrum during the event. The same region was taken from the spectrum obtained later at 15:56~UT (shown in Fig.~\ref{fig_sumer3}) to produce the reference spectrum. {\bf Bottom panel:} The spectrum during the event  is shown with a solid line, while the reference spectrum is shown with  a dotted (red in the online version of the paper) line. `W' next to the line, e.g. `N~{\sc v}~W',  denotes the blue wing of the line.}
\label{fig_sumer2}
\end{figure*}

The bright feature described above is then followed  (panel 4 in Fig.~\ref{fig_trace}) by a mostly dark (seen in absorption in EIT~195~\AA) elongated,  flux-rope-like phenomenon visible in the lower part of the dimming  at 15:34~UT. Part of the flux rope appears bright but this still does not mean that the phenomenon contains plasma at coronal temperatures because the  TRACE 195~\AA\ channel has a strong contribution from transition-region emission \citep{2006ApJS..164..202B}.  Judging from the appearance of the feature in the TRACE and CDS data, the phenomenon carries the observational description of  an eruptive filament that was also  suggested by HS.  To confirm this, we inspected the only two existing H$\alpha$ images of NOAA 08237 (Fig.~\ref{fig_fil}). The images clearly show that a filament was present in the region the day before the flare, i.e. June 12,  and part of it  disappeared after the flare took place on June 13. Ribbons are formed along the neutral line where the filament was lying a day earlier.  No other flaring activity has taken place between the time the two images were obtained, supporting the conclusion  that this filament did indeed partially erupt during the flare on June 13.

Although on relatively low cadence, the available imaging data reveal important details on the physical picture of the observed complex phenomenon. We clearly see the propagation of what appears to be  an on-disk imprint of a coronal mass ejection, i.e. a dimming region. It is apparent that the dimming/CME disintegrated with a `breaking' initiated in a small area of its leading edge. An eruptive filament is then seen moving through this `broken' region, which is then followed by a full dispersion of the dimming/CME.  To put the series of events in a clear timeline from  beginning to  end is quite challenging, again because of the limited amount of data. The CME was ejected shortly after the first flare impulsive peak as was the CW. What causes the breaking and disintegration of the CME is a  puzzle. A possible scenario is that  two consecutive  CMEs were triggered during the combined flare (i.e. two soft X-ray peaks) flare moving at different speeds, and, that  the filament is part of the second CME. The `breaking' of the first CME can then  be explained by a  CME-CME interaction,  with the second CME moving at higher speed.  The whole series of events takes place relatively low in the solar atmosphere, i.e. below one solar radius.   
The eruption of several consecutive CMEs during a `combined' flare is not an unusual phenomenon.  Most recently, \citet{2014A&A...566A.148H} reported on two consecutive CMEs ejected during a combined flare with three impulsive phases. A CME-CME interaction usually causes a change in the kinematics of one or both CMEs \citep[e.g.][]{2014ApJ...785...85T} and is  a physically sound mechanism for the observed breaking-up CME. Breaking of the CME bubble may also have occurred, caused by an impulsive CME bubble expansion that reached its maximum low in the solar atmosphere.  Impulsive cavity over-expansions also triggering EUV waves, have been already reported by  \citet{2010ApJ...724L.188P, 2010A&A...522A.100P}. The low data cadence and the limited TRACE FOV do not permit us  to perform a similar analysis.

As mentioned above, the online animation composed of a series of EIT running difference images, TRACE original, and TRACE running difference images (Fig.~\ref{fig_movie}), reveals important details on the eruptive filament. We use running difference images rather than base difference images (i.e. extracting from each image an image before the event)  to better display the propagation of the filament. In the original TRACE~195~\AA\ zoomed image, one can follow  the rise of a dark looped feature with one end  rising while moving across the solar disk. This evolution represents the typical behaviour of an asymmetric filament eruption  meaning that one leg of the filament remains anchored in the chromosphere while the other one rises and the whole filament keeps  untwisting until  the flux rope is fully `straightened' \citep[e.g.][]{1996ApL&C..34..113M, 2013ApJ...771...65J}.  The inspection of the EIT difference images shows that the filament erupted to a certain height with the `highest' position recorded in panel 12 in Fig.~\ref{fig_trace}, and then fell back to the chromosphere.  Because the eruption is non-radial, the on-disk position carries information about the `height' of the feature. 
  \citet{1999SoPh..190..467W}  reported that a CME  recorded by LASCO C2 propagating through the South Pole (i.e.  southwards from the NOAA 08237) is associated with the `EIT' wave.  The CME is first registered in the LASCO C2 at 17:06~UT at a height of 3.25R$_{\odot}$, moving with a linear speed of 179~\kms. A time--height plot (see \href{http://cdaw.gsfc.nasa.gov/CME_list/UNIVERSAL/1998_06/htpng/19980613.170605.p188s.htp.html}{the LASCO's website}) gives  a solar origin time of the phenomenon shortly before 14:00~UT, which is more than an hour before the solar flare took place. In addition, the dimming is clearly seen propagating northwards and there is no distinguishable on-disk signature of a CME propagating southwards in the EIT data.

\begin{figure*}[!ht]
\centering
\includegraphics[scale=0.9]{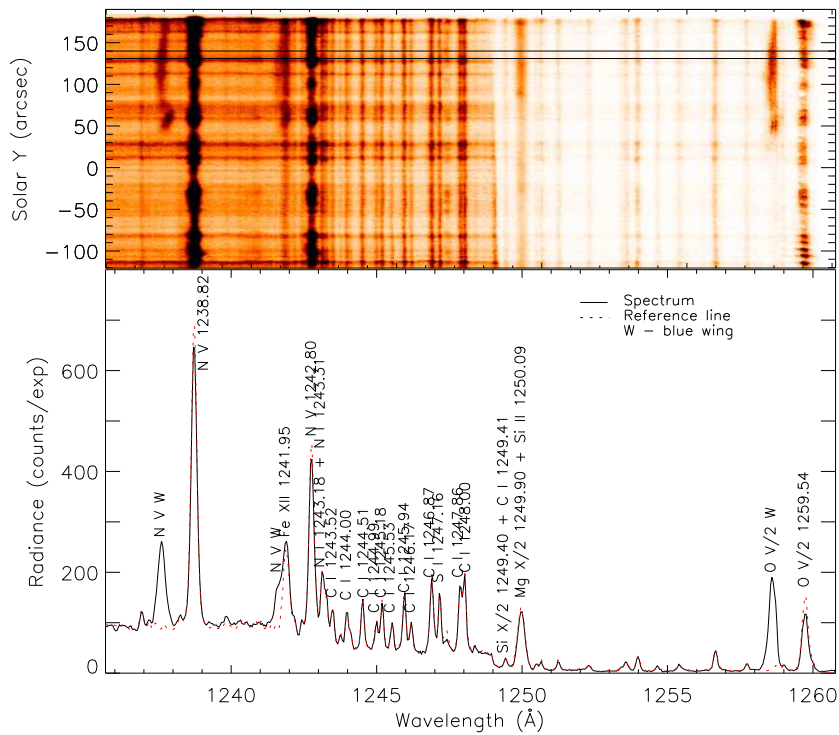}
\vspace{-1cm}
\caption{{\bf Top panel:} SUMER detector image taken at 15:56~UT. The region between the horizontal solid lines was used to produce the spectrum  during the event shown with solid line in the bottom panel. The same region was taken from the spectrum obtained earlier at 15:48~UT (shown in Fig.~\ref{fig_sumer2}) to produce the reference spectrum shown with a dotted line in the bottom panel. {\bf Bottom panel:} The spectrum during the event is shown with a solid line, while the reference spectrum is given with a dotted (red in the online version of the paper) line. `W' next to the line,  e.g. `N~{\sc v}~W',  denotes the blue wing of the line. }
\label{fig_sumer3}
\end{figure*}

\subsection{Spectral analysis}

To the  best of our  knowledge, this is the first report of SUMER observations during a coronal wave and an eruptive solar  filament. The SUMER spectrometer made four exposures, taking spectra in two wavelength bands. The first  was centred at Ne~{\sc viii} 770~\AA, while the second was in N~{\sc v} 1238~\AA.   The  registered spectral lines with a sufficient signal-to-noise are given in Table~\ref{table1}. They cover the temperature range   from 4.0 to 6.1~K  ($logT_{max}$). We should note that the Ne~{\sc viii} wavelength range only covers lines in the temperature range from  4.9 to 5.8~K ($logT_{max}$), while the N~{\sc v} spectral window contains lines with lower and higher formation temperatures (see Table~\ref{table1} for more details). 

 The filament enters the CDS FOV  after 15:34~UT, but was actually scanned by the CDS slit during the raster starting at 15:37~UT. In the CDS intensity images (Fig.~\ref{fig_vel_cds}), the twisted flux rope of the filament is clearly noticeable. The CDS data were taken in several lines, including He~{\sc i} 584.33~\AA, He~{\sc ii}  303.78~\AA, O~{\sc iv}~554.51~\AA\  (logT$_{max}$ = 5.2~K), O~{\sc v}~629.73~\AA\  (logT$_{max}$ = 5.4~K), Ne~{\sc vi}~562.71~\AA\  (logT$_{max}$ = 5.6~K), Mg~{\sc x} 609.79~\AA\  (logT$_{max}$ = 6.0~K), and Fe~{\sc xii}~364.47~\AA\  (logT$_{max}$ = 6.1~K). \citet{2003ApJ...587..429H}  analysed the three lines He~{\sc i},  O~{\sc v}, and Mg~{\sc x}  emitting at chromospheric, transition-region, and coronal temperatures. They detected  the event in all three lines, concluding that plasma at `all these temperatures is being ejected simultaneously'.  We need to make a small correction of this result and conclusion. The Mg~{\sc x}~609.79~\AA\ is blended by O~{\sc iv}~609.83~\AA\  ($logT_{max}$ = 5.2~K) and the two lines cannot be separated. In the quiet Sun and prominences the emission at 609  is dominated by  O~{\sc iv}. The blend by the  O~{\sc iv} line can be easily evaluated using the emission of  O~{\sc iv}~554.51~\AA. There are four O~{\sc iv} spectral lines  around 554~\AA,  including O~{\sc iv} 553.33~\AA, 554.076~\AA, 554.51~\AA\ and 555.26~\AA. The O ~{\sc iv} 555.26~\AA\ and  553.33~\AA\ can be excluded from the ratio calculation.  The O ~{\sc iv}  555.26~\AA\ line is too far in the red wing off 554.51~\AA, while only the red wing of O~{\sc iv} 553.33~\AA\ may have any contribution.  As the event is only registered  in the blue wing, the line will have no effect on the blue wing of 554.51~\AA. The ratio O~{\sc iv} (554.076+554.51)/609.83  derived using Chianti v.7.1.3 changes slightly with the temperature increasing from 3.65 at 100\,000~K to 4.27 at 400\,000~K, and remains close to the latter value at higher temperatures. We fitted the CDS line at 609.79~\AA\ line with a double Gaussian fit to obtain the radiance in the wing and the rest component of the line and repeated the same for the O~{\sc iv}~554.51~\AA. We used the raster starting at 15:42~UT. A small region of 5$\times$3 pixels$^2$ in the feature was used. We found a ratio for the wing of the two lines of 4.48,  suggesting that the emission in the wing of the  Mg~{\sc x} line is produced entirely by the blend of O~{\sc iv}~609.83~\AA. To summarise, the feature moving through the CDS FOV is emitting predominantly and even fully at chromospheric and transition-region temperatures, which is consistent with the temperature emission range of solar prominences. The coronal emission whether the background or related to the event,  is either negligible or above the formation temperature of Mg~{\sc x}.

The SUMER spectra were analysed by comparing them temporally and spatially with the imaging information from  TRACE, CDS-intensity and Doppler-shift rasters, and  the spectral data from CDS. The first SUMER exposure started at 15:31~UT (ended at 15:39~UT), which means that the fast wave has moved under the SUMER slit while exposing as can be seen in Fig.~\ref{fig_trace}, panels 4, 5, 6, and 7. The analysis of the first spectrum shows no detection of the wave in the temperature range from 4.9 to 5.8~K ($logT_{max}$). One of the most plausible explanations is that the wave, as known from imager data, is only detectable at temperatures higher than 1~MK, which is not recorded in the observed spectral range.  The fact that the wave is not recorded in the Mg~{\sc x} and Ne~{\sc viii} (the contribution function of Ne~{\sc viii} has a strong tail toward higher temperatures)  can also be due to the low emissivity of the wave front combined with the very long SUMER exposure time. The dimming/CME  does not make it under the SUMER slit because it starts disintegrating shortly after the beginning of the SUMER exposure.  In Fig.~\ref{fig_sumer1}, we show the spectrum taken at 15:39~UT (the second spectrum in the Ne~{\sc viii} waveband). In Figs.~\ref{fig_sumer2} and \ref{fig_sumer3}, the spectra obtained in  the N~{\sc v} wavelength range are presented.

 Next on the SUMER path is the eruptive filament. It is clearly noticeable in the bottom part of the first spectrum (Fig.~\ref{fig_sumer1}) as blue-shifted emission  in N~{\sc iv}, Ne~{\sc viii}, O~{\sc iv}, S~{\sc v}, and in some of the O~{\sc v} lines, i.e. spectral lines with transition-region  formation  temperatures  in the range 5.0 -- 5.8~K ($logT_{max}$). The filament seen in the CDS FOV (Fig.~\ref{fig_vel_cds}) moves laterally while rising in a non-radial direction in the solar atmosphere. Note that the CDS Doppler-velocity images in the O~{\sc v}~629.73~\AA\ line shown in Fig.~\ref{fig_vel_cds} are obtained by fitting a single Gaussian. This method provides relatively good Doppler-shift imaging information about the dominant shift at each spatial pixel. For each raster, the wavelength of the rest component  was obtained from the upper left part of the raster image not affected by the eruptive filament. The CDS intensity and Doppler-shift images reveal a twisted flux rope with a leading edge rising at speed above 300~\kms\ (the rasters at 15:42~UT, 15:47~UT, and 15:53~UT). The filament lower body displays at least a  3$\pi$ twist and its speed is no more than 150~\kms. These  Doppler velocities clearly show an asymmetric filament eruption.  The raster at 15:58~UT shows that the filament made a significant lateral displacement, but more importantly its speed decreased and  part of its body show velocities close to zero, meaning that the filament reached its maximum height,  decelerated and some of its material started falling back. This is consistent with the imaging information from EIT which, as we mentioned above, shows the filament stopping at a certain height. It should be noted that the CDS Doppler-shifts have errors of at least $\pm$5~\kms\ in addition to the fact that they are only relative, which adds another minimum error of $\pm$5~\kms\ to the values quoted here.

\begin{figure}[!ht]
\centering
\includegraphics[scale=0.9]{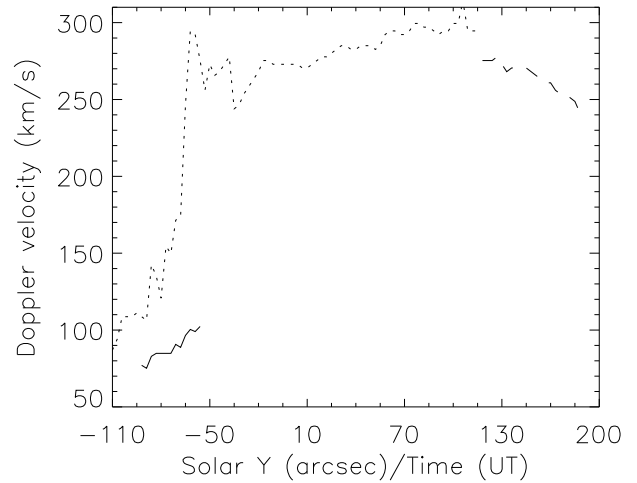}
\caption{Doppler velocity curves of the eruptive filament obtained from a double Gaussian fit in Ne~{\sc viii}~770.42~\AA\ (solid line) during the exposure from 15:39~UT to 15:47~UT, in N~{\sc v}~1238.82~\AA\ (dotted line) from 15:48~UT to 15:56~UT and from 15:56~UT to 16:03~UT (dashed line). }
\label{fig_vel_sum}
\end{figure}

 The SUMER spectrum (Fig.~\ref{fig_sumer2}) with an exposure starting at 15:48~UT shows the further apparent evolution of the feature with a sudden Doppler-shift increase, which remains relatively constant during the second (and last) exposure in the N~{\sc v} wavelength range.  Because of the lateral filament displacement and the fact that the leading edge of the filament is not yet fully up, i.e. the filament is still slightly looped during the  SUMER spectra  starting at 15:39~UT and 15:48~UT, the Doppler shifts are relatively low (up to max 100~\kms).  The filament is recorded only in spectral lines emitting at  transition region temperatures. No coronal emission was registered in the Mg~{\sc x}~624.9~\AA\ (observed in  second order) and Fe~{\sc xii}~1241.95~\AA\ lines. Once the leading fast moving edge of the filament enters the SUMER slit FOV, the measured Doppler shift seen along the SUMER slit  rises sharply. This is illustrated in Fig.~\ref{fig_vel_sum}.   Without the  imaging information from CDS and TRACE, the observed significant Doppler shift increase would have been interpreted as  acceleration of the filament. Instead, it is clear that this rising speed of the filament is due to a  faster-moving feature entering the spectrometer slit FOV.  The Doppler velocities were obtained from two, and for some positions along the slit, even up to four Gaussians, in the  Ne~{\sc viii}~770.42~\AA\ and N~{\sc v}~1238.82~\AA\ lines. Note that the Doppler-shift curves  shown in Fig.~\ref{fig_vel_sum} combine both time and space information (time: the duration of the exposure, space: the propagation of the feature along the SUMER slit).  The Doppler velocities in the rest of the lines are on the same order as the Doppler velocities in  Ne~{\sc viii} and N~{\sc v}.

\section{Summary and conclusions}

The study presented here started with the idea of providing spectroscopic insight into the coronal-wave  phenomenon by analysing SUMER observations during one such event  for the first time. During the course of the study, we extracted a wealth of information and provided for the first  time a spectroscopic description of an asymmetric eruptive filament. We found that the coronal wave registered by EIT, TRACE, SUMER, and CDS was caused by a  combined  flare with two impulsive peaks. A CME was generated during the flare and a coronal wave was observed as the CME propagated across the solar disk in a non-radial, south-north direction. We clearly identified all the features that describe the bi-modal nature of the phenomenon EUV/EIT/Coronal wave, i.e.  a fast-mode wave and CME-driven compression. A `breaking' of the CME leading edge is followed by an asymmetric filament moving `through' the partly disintegrated dimming/CME and the CME-driven compression.  We suggested that either CME-CME interaction or impulsive CME bubble over-expansion in the low solar atmosphere as a possible cause for the CME disintegration. The filament rises to a certain height and it appears to fall back to its source region. No signature of the wave was found in the spectroscopic SUMER and CDS data. An earlier detected hot-plasma emission in the Mg~{\sc x} 609.79~\AA\ by HS was found to be caused by a blend from  the O~{\sc iv}~609.83~\AA\ line.  The eruptive filament  is recorded only in a transition-region temperature range 4.9 -- 5.8~K ($logT_{max}$) and no coronal emission was detected. Thanks to the time-space combined  imaging  and spectroscopic data, a sharp increase  of the filament Doppler shift observed  in the spectroscopic data was found be caused by the movement of the fast moving leading edge of the asymmetric filament under the SUMER slit rather than being produced by a filament acceleration. We found that the twisted body of the filament rises with a speed of up to 150~\kms\ while moving laterally. The leading edge of the filament lifts up almost twice as quick with speeds of around 300~\kms.

\begin{acknowledgements}  
Research at Armagh Observatory is a grant aided by the N.~Ireland Department of Culture, Arts and 
Leisure.  MM is funded by the Leverhulme trust. We thank Richard Harrison for important discussion and an anonymous referee for insightful comments. MM thanks Kamalam Vanninathan, Astrid Veronig and Zhenghua Huang for  fruitful discussions on this manuscript. We are grateful to Vasyl Yurchyshyn for providing us with the BBSO images.  We thank STFC for the support via ST/J001082/1.
\end{acknowledgements}

\bibliographystyle{aa}

\begin{appendix}

\section{Online material}

\begin{figure*}[!h]
\centering
\includegraphics{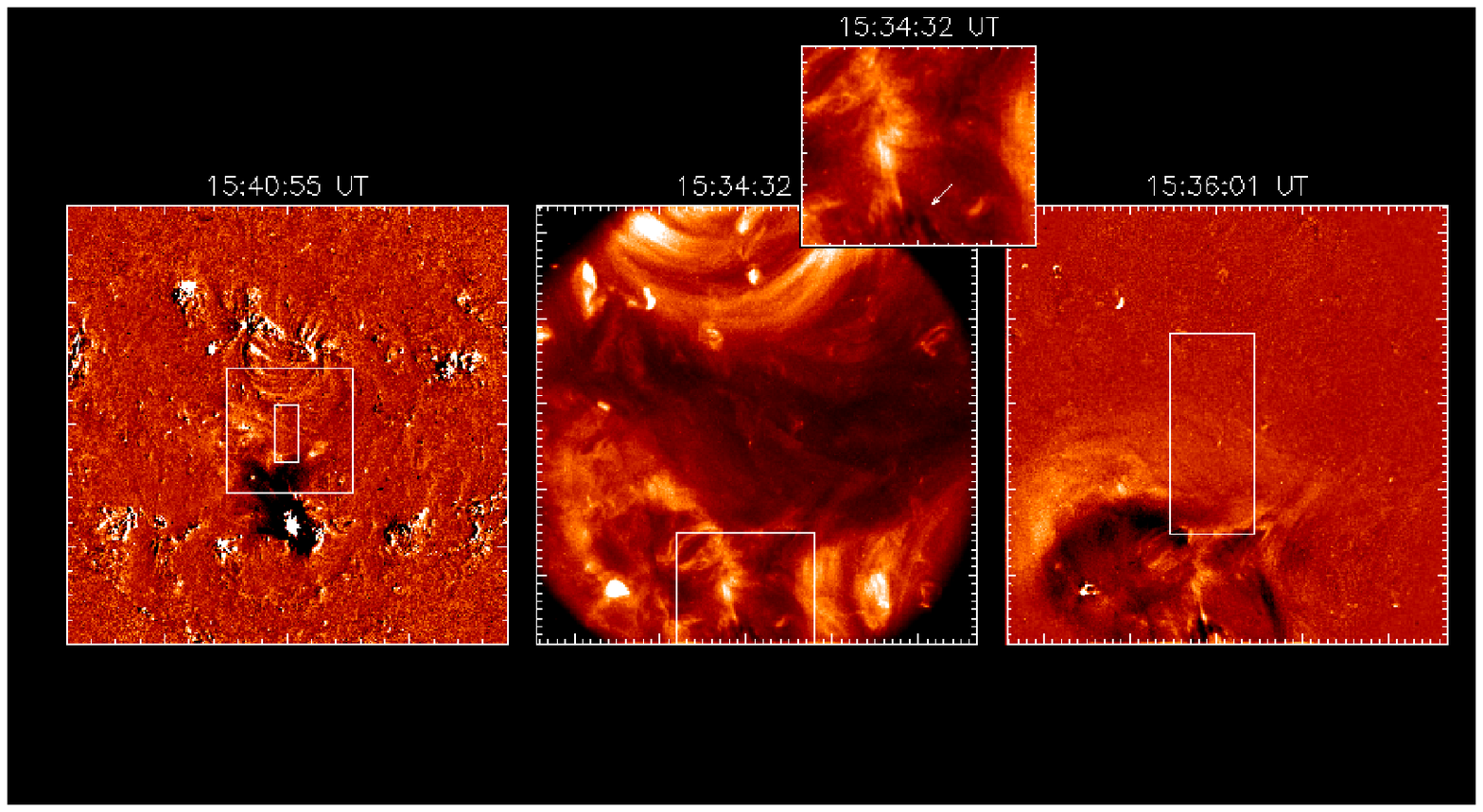}
\caption{{\bf From left to right:} EIT image running difference, TRACE original and TRACE running difference image. The square on the EIT image outlines the TRACE field-of-view while the inner rectangle is the CDS field-of-view. The TRACE running difference image is also over-plotted with CDS field-of-view. The upper small window shows a close view of the eruptive filament and is taken from the solid-line outlined region, over-plotted on the TRACE original image. }
\label{fig_movie}
\end{figure*}

\end{appendix}

\end{document}